\documentclass[nofootinbib,preprint]{revtex4-1}
\usepackage[T2A]{fontenc}
\usepackage{epsfig}

\begin{document}
\preprint{\bf PREPRINT IFJPAN-IV-2017-5}
\title{Effect of emission of extra lepton pair for precise measurement of W-boson mass
}
\author{S.~Antropov}
\email{E-mail: santrop\_2@yahoo.com}
\affiliation{\small The Henryk Niewodniczanski Institute of Nuclear Physics Polish Academy of Sciences, 31-342 Krakow, Poland}

\date{\today}

\begin{abstract}
In the paper we present results for final state emissions of lepton pairs in decays of heavy intermediate states principally of Z boson, but of some importance for the W decays as well. The presented semi-analytical calculation and {\tt PHOTOS} MC program are in numerical agreement to better than $5\%$ of pair effects. Suggestions for the future works are given.
\end{abstract}

\collaboration{Presented at "XXIII Cracow EPIPHANY Conference", 9 - 12 January 2017, Krakow, Poland. To be published in Acta Physica Polonica B}
\maketitle

\section{Introduction}

One of the purposes of LHC experiments is to improve precision of the $W$ boson mass measurement. Precision measurements of the W~boson mass rely on a precise reconstruction of momenta for the final state leptons~\cite{Aaboud:2017svj} and on comparison of W production and decay with those of Z in LHC and in particular LHC detector conditions. The QED effects of the final state radiation play an important role in such experimental studies. Final state bremsstrahlung is included in all simulation chains and should be studied together with the detector response to leptons.

\begin{figure}[htb]
\centerline{%
\includegraphics[width=16.0cm,bb=0 0 429 105]{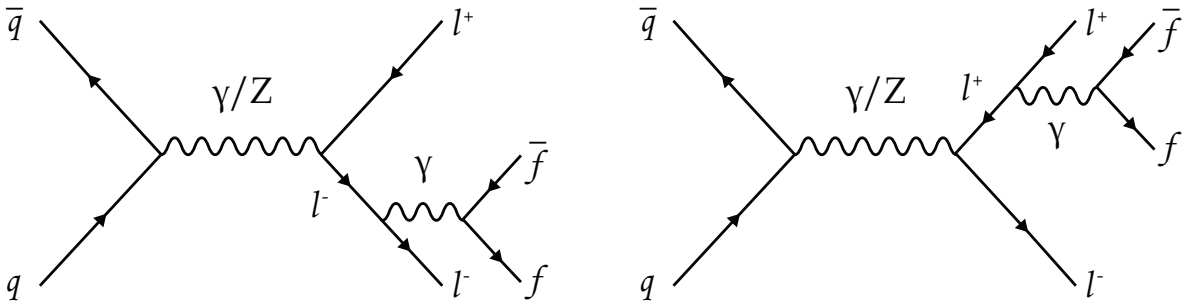}}
\caption{Real pair correction.}
\label{Fig:Antr_real_corr}
\end{figure}

In the present paper we will concentrate on the effects related to additional pair emissions in decays of heavy bosons, mainly $Z$. These effects should be included starting from the second order of QED, i.e. from the $\mathcal{O} (\alpha^2)$ corrections. The typical Feynman diagrams for pair corrections of a final states are shown in Fig.~\ref{Fig:Antr_real_corr}.

Precise calculations include radiative corrections, such corrections are usually calculated with a help of MC generators. The experimental data are compared to expectations from MC simulation. Convenient way for after-burner type algorithms is to use hard process MC generator (like {\tt PYTHIA}~\cite{Sjostrand:2007gs}) to generate series of events $pp\nolinebreak\to\nolinebreak Z/\gamma^*\to$  lepton pair and then to modify some of two lepton final states with appropriate probability to be four lepton states.

The goal of present paper is to derive analytically the correction to the $Z$ boson decay due to the emission of extra lepton pair and to compare this prediction with the previous one which is made by the {\tt PHOTOS} MC generator~\cite{Davidson:2010ew}.

\section{Description of the factorization scheme}

In what follows we implement the amplitude of emission of extra lepton pair from a final state, i.e. $Z\rightarrow\nolinebreak \ell^{+}\ell^{-}+(\ell^{+}\ell^{-})$, like the Born amplitude multiplied by a factor $\widetilde{B_f}$. The cross section of this process writes:
\begin{eqnarray}\label{antropov_sigma}
\sigma=\int d \Omega F |M_B|^2,
\end{eqnarray}
where $F$ is a factorized term\footnote{One should note that such factorization is performed in soft pair approximation, i.e. momentums of leptons before and after emission differ slightly.} for pair emission as used in ref.~\cite{Jadach:1993wk} and $\Omega$ is four body phase space, $|M_B|^2$ is square of Born level matrix element. The factorized term for pair emission $F$ is
\begin{eqnarray}\nonumber
F&=&(\frac{\alpha}{\pi})^2\frac1{\pi^2}\left( \frac{2p-a q}{a q^2 -2 p q} - \frac{2p'-a q}{a q^2 -2 p' q}\right)_\mu \left( \frac{2p-a q}{a q^2 -2 p q} - \frac{2p'-a q}{a q^2 -2 p' q}\right)_\nu\times\\\label{Antropov_Matr}
&\times&\frac{4 q_1^\mu q_2^\nu - q^2 g^{\mu\nu}}{2 q^4},
\end{eqnarray}
where $p^\mu$, $(p')^\mu$ denote the $4$-momentums of outgoing leptons, and $q_1^\mu$, $q_2^\mu$ -- of additional emitted leptons; $q^\mu\nolinebreak=\nolinebreak q_1^\mu+\nolinebreak q_2^\mu$; $q_1^2=q_2^2=\mu^2$; $p^2=(p')^2=m^2$. For the purpose of our calculations, the parameter $a$ (for more details see Ref.~\cite{Jadach:1993wk}) is set to be $0$. We will return to the question of normalization of (\ref{Antropov_Matr}) at the end of the chapter. The phase space $\Omega$ is given by
\begin{eqnarray}\nonumber
\Omega&=&\int
\frac{d^3 q_1}{2 (q_1)_0 (2\pi)^3}\cdot\frac{d^3 q_2}{2 (q_2)_0 (2\pi)^3}\cdot\frac{d^3 p}{2 p_0 (2\pi)^3}\cdot\frac{d^3 p'}{2 p'_0 (2\pi)^3} (2\pi)^4\times
\\\label{antr_phasespace}
&\times&\delta^4(R-p-p'-q_1-q_2),
\end{eqnarray}
where $R$ is the $4$-momentum of ingoing Z-boson.

After integration over angles we bring~(\ref{antropov_sigma}) to the form
\begin{eqnarray}\label{antropov_sig_fctr}
\sigma&=&\frac1{(2\pi)^6}\int\left[\frac1{(2\pi)^2}\cdot\frac{\lambda^{\frac12}(1,\frac{m^2}{s},\frac{m^2}{s})}8|M_B|^2 d cos\theta_{p} d\varphi_{p}\right]\times\widetilde{B_f},
\end{eqnarray}
where $\theta_{p}, \phi_{p}$ define orientation of $p$ in rest frame of $Z$, $s=R^2$ is square of invariant mass of lepton pair before emission of additional pair, $\lambda$-function is defined in standard way
\begin{eqnarray}
\lambda(a,b,c)=a^2+b^2+c^2-2a b-2 a c -2 b c.
\end{eqnarray}
The factor, which represents real correction due to emission of additional lepton pair, writes
\begin{eqnarray}\nonumber
\widetilde{B_f}&=&
-\frac{2}{3s}(\frac{\alpha}{\pi})^2 \lambda^{-\frac12}\left(1,\frac{m^2}{s},\frac{m^2}{s}\right)
\int \limits_{4m^2}^{(\sqrt{s}-2\mu)^2}d M^2_{LL}
\int\limits_{4\mu^2}^{(\sqrt{s}-M_{LL})^2}\frac{d M_{ll}^2}{M_{ll}^2}\times\\\label{antropov_bt_fctr}
&\times&\sqrt{1-\frac{4 \mu^2}{M_{ll}^2}} \left(1+\frac{2 \mu^2}{M_{ll}^2}\right)\times\Bigg{[}\;\frac{m^2  \sqrt{1-\frac{4 m^2}{M_{LL}^2}} \lambda^{\frac12}(s,M_{LL}^2,M_{ll}^2)}{M_{ll}^2 M_{LL}^2+\frac{m^2}{M_{LL}^2}\lambda(s,M_{LL}^2,M_{ll}^2)}+\\\nonumber
&+&\frac{M_{LL}^2-2 m^2 }{s-M_{ll}^2-M_{LL}^2}
\ln \frac{s-M_{ll}^2-M_{LL}^2-\sqrt{1-\frac{4 m^2}{M_{LL}^2}}\lambda^{\frac12}(s,M_{LL}^2,M_{ll}^2)}
{s-M_{ll}^2-M_{LL}^2+\sqrt{1-\frac{4 m^2}{M_{LL}^2}}\lambda^{\frac12}(s,M_{LL}^2,M_{ll}^2)}\Bigg{]},
\end{eqnarray}
where $M_{LL}$ is invariant mass of lepton pair after emission, $M_{ll}$ is invariant mass of additional lepton pair. The expression in square brackets in~(\ref{antropov_sig_fctr}) contains a square of the Born level matrix element and 2-body phase space (see ref.~\cite{Was:1994kg}, for example). In our semi-analytical calculation this expression corresponds to the number of events obtained by {\tt PYTHIA} for each interval of invariant masses of lepton pair.

We should stress, that in soft limit, i.e. $M_{ll}\rightarrow0$ and $M_{LL}\sim\sqrt{s}$, the factor $\widetilde{B_f}$ given by (\ref{antropov_bt_fctr}) coincide with such a factor, which is given by formula~(5) of ref.~\cite{Jadach:1993wk} for the case of emission of soft lepton pair from the initial state.

Now we return to the question of normalization of (\ref{Antropov_Matr}). The factor $\frac1{(2\pi)^6}$ in front of the~(\ref{antropov_sig_fctr}) should be omitted. This factor is a consequence of prefactor in expression for $F$ eq.~(\ref{Antropov_Matr}), where factorized part is mixed with terms which come from 2-body phase space. It is considered in ref.~{\cite{Jadach:1993wk}}.

\section{Numerical results}

With formula~(\ref{antropov_bt_fctr}) we take integrals numerically and compare results with output of {\tt PHOTOS} MC generator. For simulation by {\tt PHOTOS} and for semi-analytical calculation we first generate the sample of events from {\tt PYTHIA} with initialization as given in fig.~\ref{fig:antropov_init}. In order to complete results for {\tt PHOTOS}, its algorithm is simply applied on events generated by {\tt PYTHIA}. For calculation with formulae~(\ref{antropov_bt_fctr}) we move events, that are generated by {\tt PYTHIA}, to every possible bin with probabilities obtained from formula~(\ref{antropov_bt_fctr}).

\begin{figure}[htp!]
\centering
\begin{minipage}[h]{0.5\linewidth}
\footnotesize{\linespread{0.956}
\begin{verbatim}
WeakSingleBoson:ffbar2gmZ = on
23:onMode = off
23:onIfAny = 11
23:mMin    = 10.0
23:mMax    = 200.0
HadronLevel:Hadronize = off
SpaceShower:QEDshowerByL = off
SpaceShower:QEDshowerByQ = off
PartonLevel:ISR = off
PartonLevel:FSR = off
Beams:idA =   2212
Beams:idB =   2212
Beams:eCM =  14000.0
\end{verbatim}
}
\end{minipage}
\vskip 0.3cm \caption[] {Initialization parameters for {\tt PYTHIA}.
\label{fig:antropov_init}}
\end{figure}

\begin{figure}[htp!]
\centerline{%
\includegraphics[width=8.0cm]{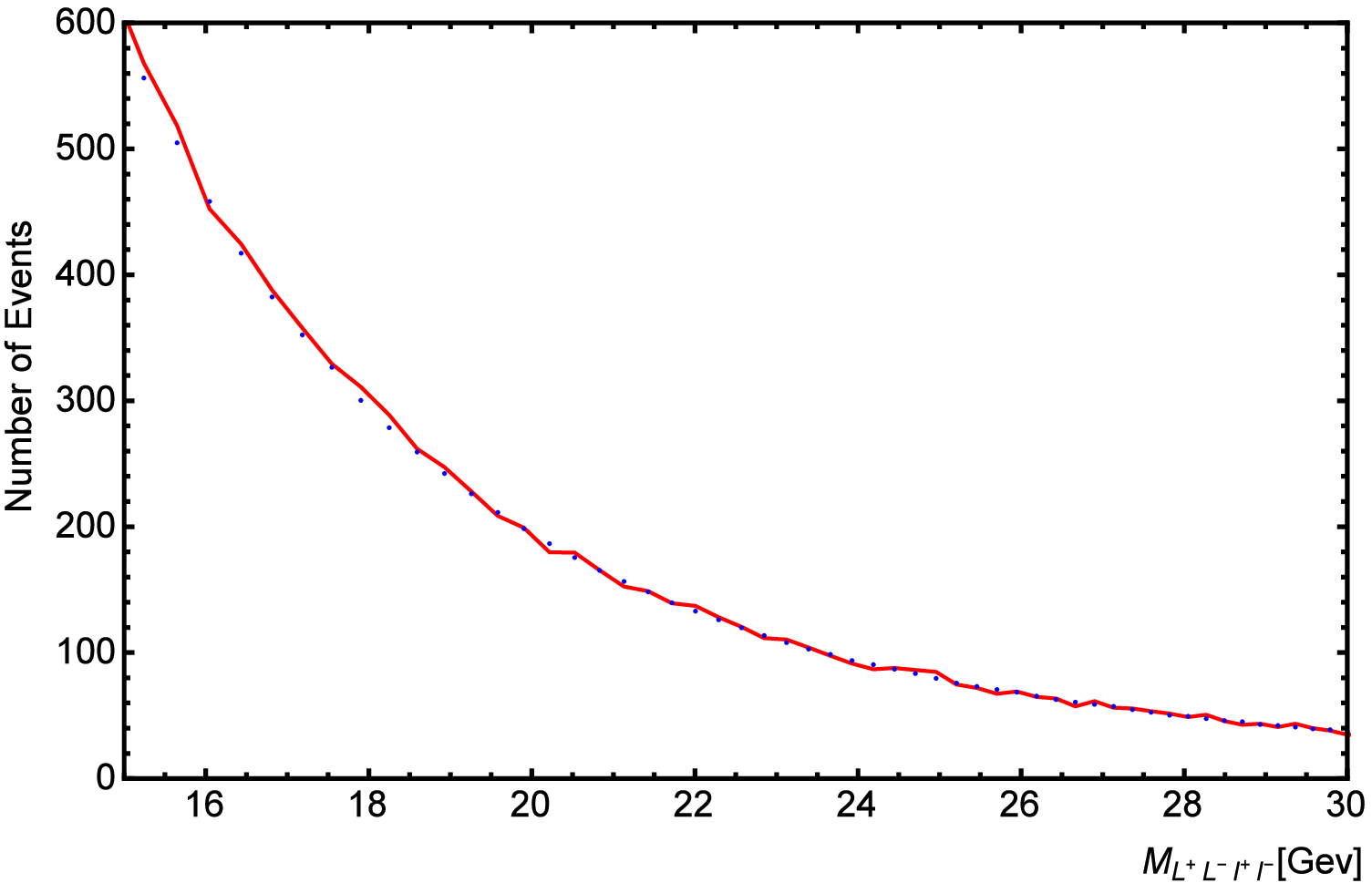}
\includegraphics[width=8.0cm]{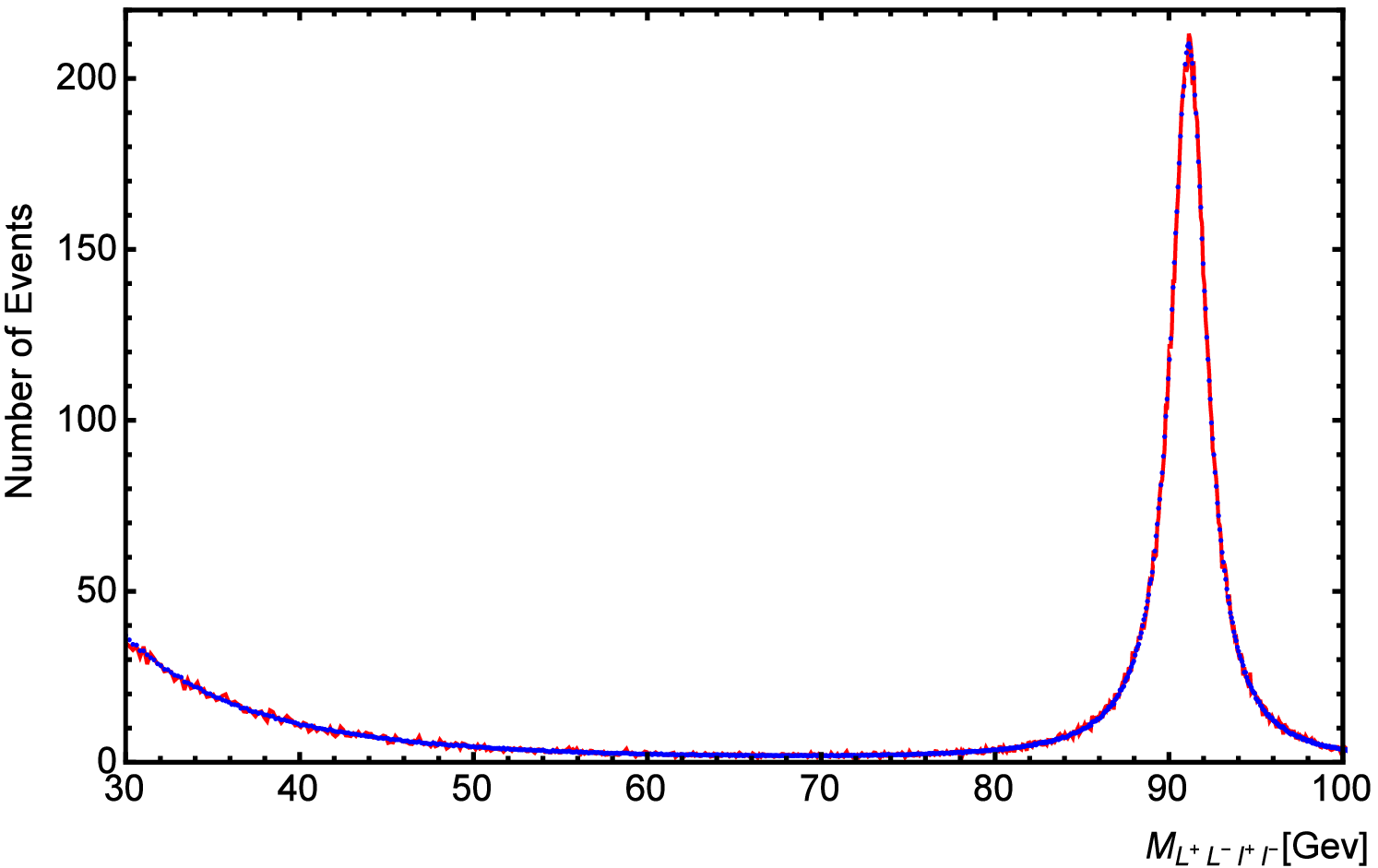}
}
\caption[] {Comparison of {\tt PHOTOS} and semi-analitical calculations for the process $p p\rightarrow\nolinebreak Z\nolinebreak\rightarrow\nolinebreak e^{+}e^{-}\nolinebreak(e^{+}e^{-})$. Solid line represents data by {\tt PYTHIA}$\times${\tt PHOTOS}. Points represent simulation by {\tt PYTHIA}, modified with formula~(\ref{antropov_bt_fctr}).
\label{fig:antropov_ee}}
\end{figure}

In fig.~\ref{fig:antropov_ee} distribution of the number of events of Z production and decay as a function of invariant mass of four leptons is presented. Solid line represents data by {\tt PYTHIA}$\times${\tt PHOTOS}. Points correspond to semi-analitical evaluation.

\begin{figure}[htp!]
\centerline{%
\includegraphics[width=8.0cm]{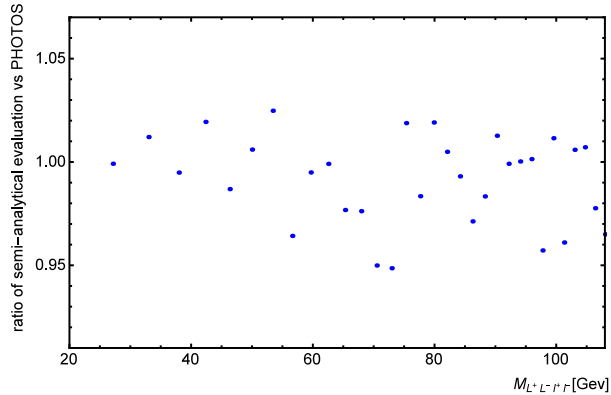}
}
\caption[] {Number of events from {\tt PYTHIA} multiplied by a factor resulting from formula~(\ref{antropov_bt_fctr}) divided by number of events from {\tt PYTHIA}$\times${\tt PHOTOS}.
\label{fig:antropov_ratio}}
\end{figure}

In fig.~\ref{fig:antropov_ratio} ratio of the two last distribution as a function of invariant mass of four leptons is presented.

Analyzing the figs.~\ref{fig:antropov_ee} and~\ref{fig:antropov_ratio} we can conclude, that {\tt PHOTOS} is in agreement with analytical calculation fairly well. Numerical precision of agreement is better than $5\%$ of the pair effect. Estimation is limited by the numerical calculation and CPU time. It can be improved rather easily.

\section{Conclusions}

Computation of distribution of the number of emissions of additional electron-positron pair in the process $p p\rightarrow\nolinebreak Z\rightarrow\nolinebreak e^{+}e^{-}$ on the invariant mass of two electron-positron pairs by {\tt PHOTOS} is in good agreement with computation based on analytic formula. Our calculation is based on an extension of previous calculation~\cite{Jadach:1993wk}, where soft approximation for emission of pairs was used. Since rigorous expression~(\ref{antr_phasespace}) for 4 body phase space is implemented in {\tt PHOTOS}, it can be developed to operate full matrix element for additional lepton pair emission similarly to modifications that are used in refs.~\cite{Nanava:2009vg,Golonka:2006tw} for photon emission.

Semi-analytical and {\tt PHOTOS} approaches allow to investigate processes with any type of emitting and emitted lepton pairs, i.e. processes $p p\rightarrow Z\rightarrow e^{+}e^{-}(e^{+}e^{-},\mu^{+}\mu^{-})$ and $p p\rightarrow Z\rightarrow \mu^{+}\mu^{-}(e^{+}e^{-},\mu^{+}\mu^{-})$. Such investigation can be a further step of our research.

Presented test complements {\tt PHOTOS}  comparison to results for 4 fermion final states obtained from {\tt KORALW} Monte Carlo \cite{Jadach:1998gi} for Z~boson decay, where extremely narrow width of intermediate Z boson is used to block emission of pairs from initial state, but preserve gauge invariance. Further tests are reported in ref.~\cite{Davidson:2010ew}.

\vspace*{0.5cm}
\acknowledgements
The author is very grateful to Prof.~Dr.~Z.~Was for supervision of this research and to Prof.~Dr.~A.~Arbuzov, Prof.~Dr.~M.~Skrzypek and Prof.~Dr.~R.~Sadykov for considerations and advise.



\begin{thebibliography}{99}

\bibitem{Aaboud:2017svj}
  M.~Aaboud {\it et al.} [ATLAS Collaboration],
  arXiv:1701.07240 [hep-ex].

\bibitem{Sjostrand:2007gs}
  T.~Sjostrand, S.~Mrenna and P.~Z.~Skands,
  Comput.\ Phys.\ Commun.\  {\bf 178}, 852 (2008)
  doi:10.1016/j.cpc.2008.01.036
  [arXiv:0710.3820 [hep-ph]].



\bibitem{Davidson:2010ew}
N.~Davidson, T.~Przedzinski and Z.~Was,
  Comput.\ Phys.\ Commun.\  {\bf 199}, 86 (2016)
  doi:10.1016/j.cpc.2015.09.013
  [arXiv:1011.0937 [hep-ph]].

\bibitem{Jadach:1993wk}
  S.~Jadach, M.~Skrzypek and B.~F.~L.~Ward,
  Phys.\ Rev.\ D {\bf 49}, 1178 (1994).
  doi:10.1103/PhysRevD.49.1178.

\bibitem{Was:1994kg}
  Z.~Was, 1993 European School of High-Energy Physics, Zakopane,
  Poland, 12-25 Sep 1993: Proceedings,
  CERN-TH-7154-94, C93-09-12, [https://inspirehep.net/record/372028/].

\bibitem{Nanava:2009vg}
  G.~Nanava, Q.~Xu and Z.~Was,
  Eur.\ Phys.\ J.\ C {\bf 70}, 673 (2010)
  doi:10.1140/epjc/s10052-010-1454-8
  [arXiv:0906.4052 [hep-ph]].


\bibitem{Golonka:2006tw}
  P.~Golonka and Z.~Was,
  Eur.\ Phys.\ J.\ C {\bf 50}, 53 (2007)
  doi:10.1140/epjc/s10052-006-0205-3
  [hep-ph/0604232].


\bibitem{Jadach:1998gi}
  S.~Jadach, W.~Placzek, M.~Skrzypek, B.~F.~L.~Ward and Z.~Was,
  Comput.\ Phys.\ Commun.\  {\bf 119}, 272 (1999)
  doi:10.1016/S0010-4655(99)00219-2
  [hep-ph/9906277].

\end{thebibliography}
\end{document}